# Names without information: Attention allocation and rent transfer in a zero-fundamental token market

Dingding Cao[a], Han Wang[a], Yujing Zhong[a], Xian Pan[b], Rizwan Akhtar[c], Wei Yang[a,*]

[a] *School of Big Data, Baoshan University, Baoshan 678000, Yunnan, China*

[b] *School of Business, Guangzhou College of Technology and Business, Guangzhou 510850, China*

[c] *School of Computing Sciences, Pak-Austria Fachhochschule: Institute of Applied Sciences and Technology (PAF-IAST), Haripur, Pakistan*

* Corresponding author. E-mail: bsxy10202@bsc.edu.cn

## Abstract

Asset names are associated with investor trading and asset prices, but where names and issuer quality are formed jointly, information, preference and attention-coordination explanations are difficult to distinguish. We study a token launchpad on which tokens issued under the default template use the same contract code, have a fixed supply and carry no cash flows, while names can be registered at almost no cost, cannot be verified and may be reused. Using all 458,174 default-template tokens launched from February to June 2026 and a sample split by creator entity fixed in advance, we find that the share of tokens attracting an outside buyer rises from 34.9% to 52.8% across name-appeal deciles, but within creator and creation minute the effect of appeal is small and does not replicate in the validation sample. Conditional on early capital inflow and creator fixed effects, the residual effects of names on migration, peak market capitalization and post-peak drawdown are statistically equivalent to zero, and buyers who follow appealing names earn no economically meaningful premium. The same name attracts less capital with each reuse, names whose first token draws more capital are copied sooner, and competition over names mainly transfers wealth among buyers.

## 1. Introduction

Firms with fluent names have broader investor bases and higher valuations [1], newly listed stocks with pronounceable tickers perform better on their first day of trading [2], and stocks whose tickers are English words trade more actively [3]. Similar associations appear in stock returns [4] and initial coin offerings [5]. Three explanations have been offered. The information explanation holds that a name reflects issuer quality, and firms with fluent names are indeed more profitable [4]. The preference explanation holds that investors favor easily processed names for their own sake [2], and preregistered experiments find a significant non-deliberate fluency effect, whereas a deliberate reading of names as quality signals can be switched on or off in boundary cases [6]. The coordination explanation holds that a name serves as a focal point for attention [7], which draws capital because buying decisions are constrained by attention [8].

In markets with fundamentals these explanations are hard to separate, because a name and its issuer's quality are formed together and a capable team and a merely noticeable name leave similar traces in prices.

Token launchpads provide a setting in which the three explanations make different predictions. On the platform studied here, tokens issued under the default template use the same contract code, have a fixed supply and carry no cash-flow rights, so a name cannot reflect cash flows, although it may still be correlated with characteristics of its creator. Names are almost free to register, their content cannot be verified, and the interests of issuers and buyers conflict. Under such a conflict, costless messages convey at most coarse information, and less of it the larger the conflict [9]. This does not rule out partial transmission, so whether names carry information remains an empirical question. Different creators can also reuse the same name, which holds its content fixed while the context of its use changes.

We use all 458,174 tokens launched under the default template from February to June 2026, drawn from a census of 513,647 launches, and measure name appeal with the three-layer framework of a companion study [10]. The paper tests predictions that separate the sources of name effects and measures the wealth transfers that competition over names generates. It also adds to research on investor attention in crypto-asset markets [11,12], attention-driven herding [13], salience [14], trading manipulation [15,16] and memecoin risk [17].

## 2. Research design

### 2.1 Institutional setting and data

We study the four.meme launchpad on BNB Chain, whose mechanism resembles that of pump.fun on Solana. Each token has a fixed supply of one billion units and initially trades on a deterministic bonding curve, and once the curve has raised its funding target, liquidity migrates to a public decentralized exchange. The platform's contract records a launch template for every token. We restrict the analysis to the 458,174 tokens (89.20%) of the 513,647 launched from 1 February to 30 June 2026 that use the default template, which levies no transaction tax and makes no distributions to holders. All 2,000 default-template tokens in a random sample have identical runtime bytecode. The excluded launches comprise 38,652 tax-template tokens, which may levy a transaction tax and distribute part of it to holders, and 16,821 exclusive (X Mode) tokens, which require a separate purchase method.

Each token is observed for 30 days after creation. The 44,271 tokens (9.66%) that never traded are retained to avoid survivorship bias, and 99.61% never migrated. On-chain events come from Dune Analytics and market data from the public OKX interface, covering 44,645,911 trades, and each token's template is read from the platform's contract through a public node. Following the companion study, we call all tokens that share the same normalized name a name lode [10].

The sample is split by creator entity so that tokens of one operator never enter both subsamples. A union-find procedure on first-funding addresses merges the 35,677 creator addresses of the census into 9,139 entities, which are stratified by the week of their first launch and by launch volume and randomly assigned before any estimation, giving 235,795 exploration and 222,379 validation tokens.

### 2.2 Variables

In the companion framework [10], the form layer measures processing fluency, the reference layer whether a name refers to objects outside the corpus, and the relational layer the context of its appearance, including crowding by recent similar names and use order within the lode. Each layer's features are combined into a sub-index by inverse-covariance weighting [18], and the appeal index is the standardized linear prediction of early inflow from the three sub-indices, with weights estimated on the exploration sample and then frozen. Because the index is fitted to early inflow in the exploration sample, exploration estimates of attention contain an in-sample fitting component, whereas validation estimates do not.

Because wash trading is documented on centralized crypto exchanges [15,16], purchases within 60 minutes of creation by the creator's address, or by any address that also sells the token in that window, are counted as wash trades and all other purchases as outside capital. The rule cannot detect coordination among unlinked addresses, so outside buyers are buyers identified by this rule rather than verified independent investors. Among exploration tokens with a purchase in the window, 50.17% attract no outside capital, so the main attention measure is whether a token attracts an outside buyer. Early inflow, the inverse hyperbolic sine of the BNB amount bought on the bonding curve within 60 minutes, controls for the attention a token attracts, with other quote assets converted at the daily BNB price. Market outcomes are migration, peak market capitalization, post-peak drawdown and survival time, and the buyer-level outcome is the holding-period return of outside buyers.

### 2.3 Testable predictions and testing procedure

Table 1 lists the predictions of the three explanations under two maintained assumptions. We assume that information in a name is not fully priced by the first buyers, in line with evidence that some investors neglect the information in fluent names [4], and that preferences are static, so that a name's utility does not depend on how often it has been used. Under these assumptions, all three explanations imply that names raise the probability of attracting an outside buyer. The information explanation further predicts that names still predict market outcomes once early inflow is controlled for and that buyers who follow appealing names earn higher returns. A static preference implies that appeal does not change with use order, whereas under the coordination explanation as formulated here, appeal weakens with repeated use and names that draw capital are copied quickly.

**Table 1 Testable predictions of the three explanations under the maintained assumptions**

| Test | Information | Preference | Coordination |
|---|---|---|---|
| Name appeal raises the probability of attracting an outside buyer | Yes | Yes | Yes |
| Residual effect of names on market outcomes, conditional on early inflow | Non-zero | Zero | Zero |
| Returns of outside buyers who follow appealing names | Higher | Not higher | Not higher |
| Appeal of the same name under repeated use | — | Unchanged | Declines with use order |
| Speed at which other creators copy a name | — | — | Fast, and faster when the first token draws more capital |

Notes: The predictions rest on two maintained assumptions, that information in a name is not fully priced by the first buyers and that preferences do not depend on a name's use order (Section 2.3). The last two rows are the empirical implications of the coordination explanation examined here. A dash indicates that the explanation makes no clear prediction for the test. Outside buyers are buyers identified by the rule in Section 2.2. Early inflow is the amount bought on the bonding curve in the first 60 minutes after a token is created.

All variable definitions, specifications and decision rules were frozen on the exploration sample before the validation sample was used. The confirmatory tests were first run on data that included all launch templates and recorded the early inflow of stablecoin-quoted tokens in US dollars. After restricting the sample to the default template and converting early inflow to BNB, we recomputed the exploration estimates and frozen constants and reran each validation test once, and both runs are reported in the replication package. Null results are assessed with equivalence tests [19]. Each regression includes a positive control expected to have a strong effect, and an effect is judged equivalent to zero if the 90% confidence interval of the appeal coefficient lies within a pre-specified bound of 5% of the positive control's standardized effect, which we regard as economically negligible. Standard errors are clustered by creator entity.

## 3. Results

### 3.1 Names and attention allocation

Across appeal deciles, the share of exploration tokens that attract an outside buyer within 60 minutes rises from 34.9% to 52.8% (Figure 1a). Column (1) of Table 2 adds creation-minute and creator fixed effects, which absorb common market conditions and time-invariant differences between operators. A one-standard-deviation increase in appeal then raises this probability by 0.29 percentage points ($p$ = 0.083), against a share of 45.3%, and the validation estimate is 0.15 percentage points ($p$ = 0.382). With early inflow as the dependent variable, the appeal coefficient is 0.0267 ($p$ = 0.017) with creation-minute fixed effects and −0.0049 ($p$ = 0.094) once creator fixed effects are added. The association between names and attention thus arises mainly between creators, and within a creator it is small and not confirmed out of sample.

**Table 2 Name appeal and early attention**

| | (1) | (2) | (3) | (4) |
|---|---|---|---|---|
| Dependent variable | Outside buyer | Outside buyer | Early inflow | Early inflow |
| Sample | Exploration | Validation | Exploration | Exploration |
| Name appeal index | 0.0029* | 0.0015 | 0.0267** | −0.0049* |
| | (0.0017) | (0.0017) | (0.0111) | (0.0029) |
| Creation-minute fixed effects | Yes | Yes | Yes | Yes |
| Creator fixed effects | Yes | Yes | No | Yes |
| Observations | 182,447 | 170,137 | 191,159 | 182,447 |
| Mean of dependent variable | 0.453 | 0.446 | 0.830 | 0.816 |

Notes: The unit of observation is the token, and all tokens use the default template. The analysis sample excludes the 38,652 tokens (7.53%) launched under tax templates, 6,325 of which distribute part of the transaction tax to holders, and the 16,821 exclusive (X Mode) tokens (3.27%), whose funding target is 50 BNB. The funding target read from the platform's contract is 18 BNB for all 422,202 default-template tokens quoted in BNB. Of the default-template tokens, 13,188 (2.88%) use the optional anti-sniper fee schedule, under which the trading fee is higher in the first blocks after creation, and are retained because the schedule gives holders no cash flows. Outside buyer equals 1 if the token attracts at least one outside buyer within 60 minutes of creation, with outside buyers identified by the rule in Section 2.2. Early inflow is the inverse hyperbolic sine of the amount bought on the bonding curve (in BNB) within 60 minutes, with amounts in other quote assets converted at the daily BNB price. The name appeal index is standardized on the exploration sample, and its weights are estimated on the exploration sample and then frozen. The exploration sample contains 234,870 tokens with a non-empty name. The number of observations is smaller because fixed-effect groups with a single observation provide no within-group comparison and are dropped. Means of the dependent variables are computed over the estimation sample of each column. Standard errors clustered by creator entity are in parentheses. *, ** and *** denote significance at the 10%, 5% and 1% levels, respectively.

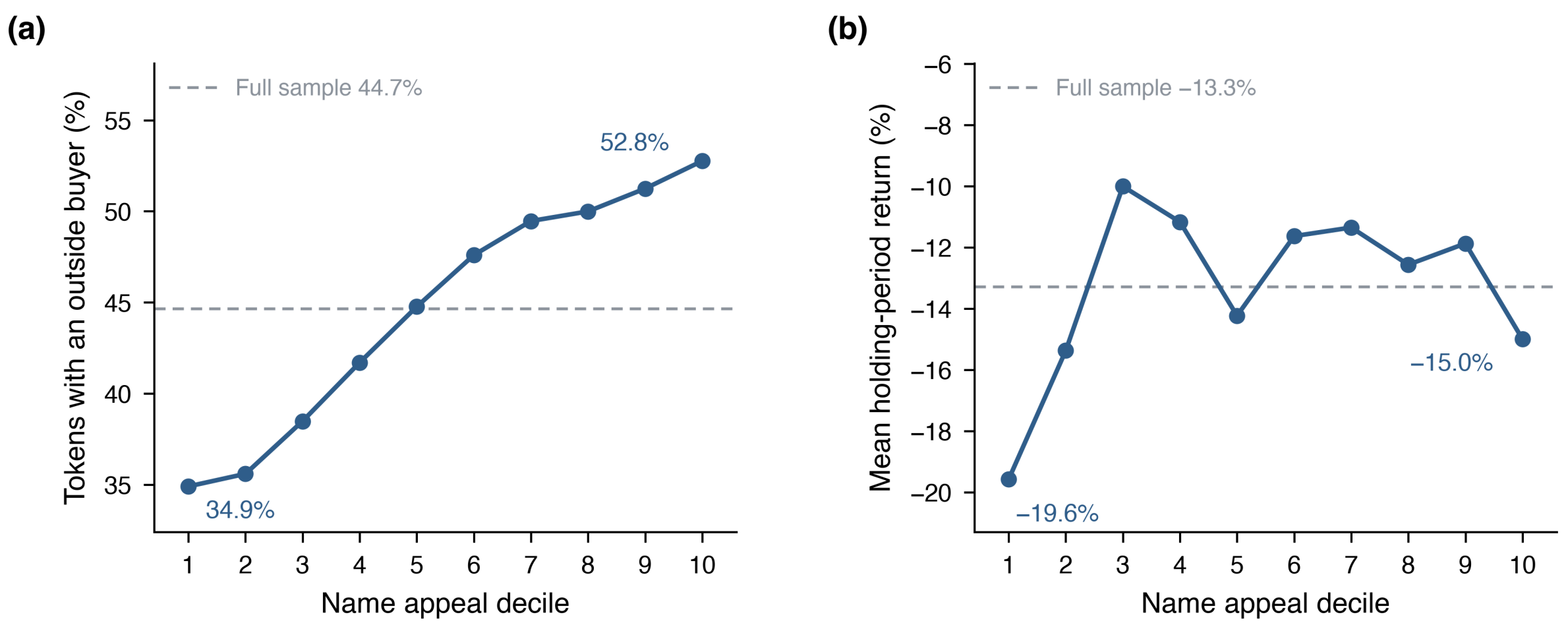

**Figure 1.** Name appeal, outside-buyer entry and returns. (a) Share of tokens that attract an outside buyer within 60 minutes of creation, by name appeal decile, for the 234,870 default-template exploration tokens with a non-empty

name. The dashed line marks the full-sample share of 44.7%. (b) Mean 30-day holding-period return of outside buyers, by name appeal decile, for 4,591,951 outside buyer–token pairs in the exploration sample. The dashed line marks the full-sample mean of −13.3%. Decile cut-offs are set on the exploration sample.

### 3.2 Residual effects conditional on attention

Table 3 adds early inflow to the outcome regressions, so the appeal coefficient captures residual effects beyond attention. Early inflow strongly predicts all four outcomes, with absolute *t*-statistics between 15.8 and 158.6. For migration, peak market capitalization and post-peak drawdown, the 90% confidence intervals lie within the equivalence bounds in both subsamples and admit at most 2.8% of the standardized effect of early inflow, although the exploration coefficient for peak market capitalization is significantly negative ($p < 0.001$). The hazard model for survival time, which as pre-specified includes early inflow and survival-day dummies but no fixed effects, yields a significantly positive coefficient in both subsamples, raising the hazard of ceasing to trade by 4.8% and 4.7% per standard deviation of appeal. In the exploration sample, the linear regression with creation-minute and creator fixed effects for which the bound was set gives a coefficient inside the bound (Table 3 notes), so the survival result depends on the specification and we make no equivalence judgment.

**Table 3 Residual effects of name appeal on market outcomes conditional on early inflow**

| | Migration | Peak market cap | Post-peak drawdown | Survival time |
|---|---|---|---|---|
| Estimator | Linear probability | OLS | OLS | Discrete-time hazard |
| **Panel A. Exploration sample** | | | | |
| Name appeal index | −0.00027 | −0.00208*** | −0.00025 | 0.04702*** |
| | (0.00017) | (0.00055) | (0.00041) | (0.00444) |
| Early inflow | 0.02161*** | 0.31287*** | 0.16732*** | −0.40415*** |
| | (0.00137) | (0.00512) | (0.00106) | (0.00887) |
| 90% CI of appeal coefficient | [−0.00054, 0.00001] | [−0.00299, −0.00117] | [−0.00093, 0.00043] | — |
| Corrected appeal coefficient | 0.00145*** | — | — | — |
| | (0.00042) | | | |
| Observations | 182,447 | 159,721 | 67,439 | 499,952 |
| **Panel B. Validation sample** | | | | |
| Name appeal index | −0.00024 | −0.00092 | 0.00042 | 0.04623*** |
| | (0.00017) | (0.00071) | (0.00039) | (0.00752) |
| 90% CI of appeal coefficient | [−0.00052, 0.00004] | [−0.00209, 0.00025] | [−0.00022, 0.00105] | — |
| Observations | 170,137 | 152,304 | 60,466 | 448,010 |
| Equivalence bound | ±0.000965 | ±0.0141 | ±0.0088 | — |
| Equivalent to zero in both subsamples | Yes | Yes | Yes | — |
| Creation-minute and creator fixed effects | Yes | Yes | Yes | No |

Notes: Migration equals 1 if the token migrates. Peak market cap is the inverse hyperbolic sine of the peak market capitalization (in US dollars) within the 30-day window and is defined only for tokens that trade within the window. Post-peak drawdown is the decline of the lowest post-peak price relative to the peak price and is defined only for tokens that still trade after the peak. Survival time is the time from creation to the last trade, with tokens still trading at the end of the window right-censored at 30 days. It is estimated with a discrete-time hazard model with a complementary log-log link on token-day observations that, as pre-specified, includes survival-day dummies (pooled from day 15 onward in the validation sample) but no fixed effects, and a positive coefficient indicates a higher hazard of ceasing to trade. All regressions control for early inflow, and equivalence judgments refer to estimates without correction for measurement error. The equivalence bound is 5% of the standardized effect of early inflow in the same regression, set on the exploration sample and applied to the validation sample. The bound for survival time was set for a linear regression of the inverse hyperbolic sine of survival time with creation-minute and creator fixed effects, which is not on the scale of the hazard model, so no equivalence judgment is made. In that linear regression, the exploration appeal coefficient is 0.0003, with a 90% confidence interval of [−0.0061, 0.0067] and a bound of ±0.0347. The corrected appeal coefficient adjusts for measurement error by using two indices, built from randomly split halves of the features of each layer, as instruments for each other and, as pre-specified, is estimated only for migration and with creation-minute fixed effects only. In that specification, the uncorrected coefficient is −0.00107*** (0.00024). The first-stage $F$-statistic is 221.6, and the correlation between the two indices is 0.22. In the pre-specified specification curve, which varies the outcome, the fixed effects, the sample restriction and the language of the name, 21.9% of the 128 specifications that use the appeal index and control for early inflow yield an appeal coefficient significant at the 5% level (4.7% positive and 17.2% negative). Standard errors clustered by creator entity are in parentheses. *, ** and *** denote significance at the 10%, 5% and 1% levels, respectively.

A null result may also reflect measurement error in the appeal index. We split the features of each layer at random into two halves, aggregate each half into an index with the frozen weights, and use the two indices as instruments for each other [20]. The correction assumes that both halves measure the same latent appeal with uncorrelated errors, but the halves correlate at only 0.22, so they may not be parallel measures and the direction of any remaining bias is unknown. Under this assumption, and with creation-minute fixed effects only as pre-specified, the residual migration coefficient is 0.00145 ($p < 0.001$), which exceeds the bound of ±0.000965 (Table 3, Panel A). Without the correction, the coefficient in this specification is −0.00107 ($p < 0.001$), also outside the bound, so the equivalence of the migration residual holds only when creator fixed effects are included.

Table 4 tests whether outside buyers who follow appealing names earn higher returns. With entry-minute fixed effects and controls for entry order, trading experience and purchase cost, the appeal coefficient is −0.0049 ($p = 0.096$) in the exploration sample, −0.0025 ($p = 0.199$) with buyer fixed effects and 0.0003 ($p = 0.152$) after the correction for measurement error. The validation coefficient is 0.0072 ($p = 0.003$), significantly positive but with a 90% confidence interval entirely within the equivalence bound of ±0.0179, which admits at most 3.1% of the effect of entry order. Mean holding-period returns are negative in every appeal decile and not monotone in appeal, at −19.6% in the lowest decile, −10.0% in decile 3 and −15.0% in the highest (Figure 1b). By comparison, moving from the earliest to the latest entry lowers the return by 35.9 percentage points. The return premium predicted by the information explanation is either absent or economically negligible.

**Table 4  Name appeal and holding-period returns of outside buyers**

| | (1) | (2) | (3) | (4) |
|---|---|---|---|---|
| Sample | Exploration | Exploration | Exploration | Validation |
| Name appeal index | −0.0052 | −0.0049* | −0.0025 | 0.0072*** |
| | (0.0034) | (0.0029) | (0.0019) | (0.0024) |
| Entry order | | −0.3589*** | −0.3814*** | −0.4688*** |
| | | (0.0111) | (0.0084) | (0.0093) |
| Corrected appeal coefficient | | 0.0003 | | |
| | | (0.0002) | | |
| Trading experience and purchase cost | No | Yes | Yes | Yes |
| Entry-minute fixed effects | Yes | Yes | Yes | Yes |
| Buyer fixed effects | No | No | Yes | No |
| 90% CI of appeal coefficient | | | | [0.0033, 0.0112] |
| Observations | 4,579,735 | 4,579,735 | 4,254,797 | 4,162,767 |

Notes: The unit of observation is an outside buyer–token pair. The dependent variable is the 30-day holding-period return, equal to sale proceeds plus the value of the remaining position at the end-of-window price, minus the purchase cost, divided by the purchase cost. It is winsorized at the 1st and 99th percentiles of the exploration sample, and the validation sample uses the exploration bounds. Entry order is the buyer's rank of entry among all buyers of the token, scaled so that 0 is the earliest and 1 the latest entry. Trading experience and purchase cost are the number of tokens the buyer has bought before and the cost of the current purchase, both in inverse hyperbolic sine. All regressions include a censoring indicator that equals 1 if the token is still trading and the buyer still holds a position at the end of the window. The equivalence bound for column (4) is 5% of the absolute exploration coefficient on entry order, that is, ±0.0179, and the 90% confidence interval lies entirely within this bound. The corrected appeal coefficient adjusts for measurement error as in Table 3 and, as pre-specified, is estimated with the controls of column (2) and fixed effects for the token's creation minute instead of the entry minute. Standard errors clustered by the entity of the token's creator are in parentheses. *, ** and *** denote significance at the 10%, 5% and 1% levels, respectively.

### 3.3 Context dependence and the destination of rents

In clean repeated lodes of the exploration sample, mean early inflow is 1.35 at a name's first use, 0.83 at the second and 0.62 at the twentieth (Figure 2). With lode fixed effects absorbing the appeal of the name itself and creation-minute fixed effects controlling for market conditions, the use-order coefficient is −0.456, and −0.449 in the validation sample (Table 5, Panel A). When imitation is separated from batch issuance by the time between consecutive uses and the difference between names, the two kinds of repetition decline at rates that do not differ significantly. Appeal thus falls with use order while the name stays the same. This does not support a static preference and is consistent with a focal point being used up with each use, although preferences that change with repeated exposure or with the investor base cannot be excluded.

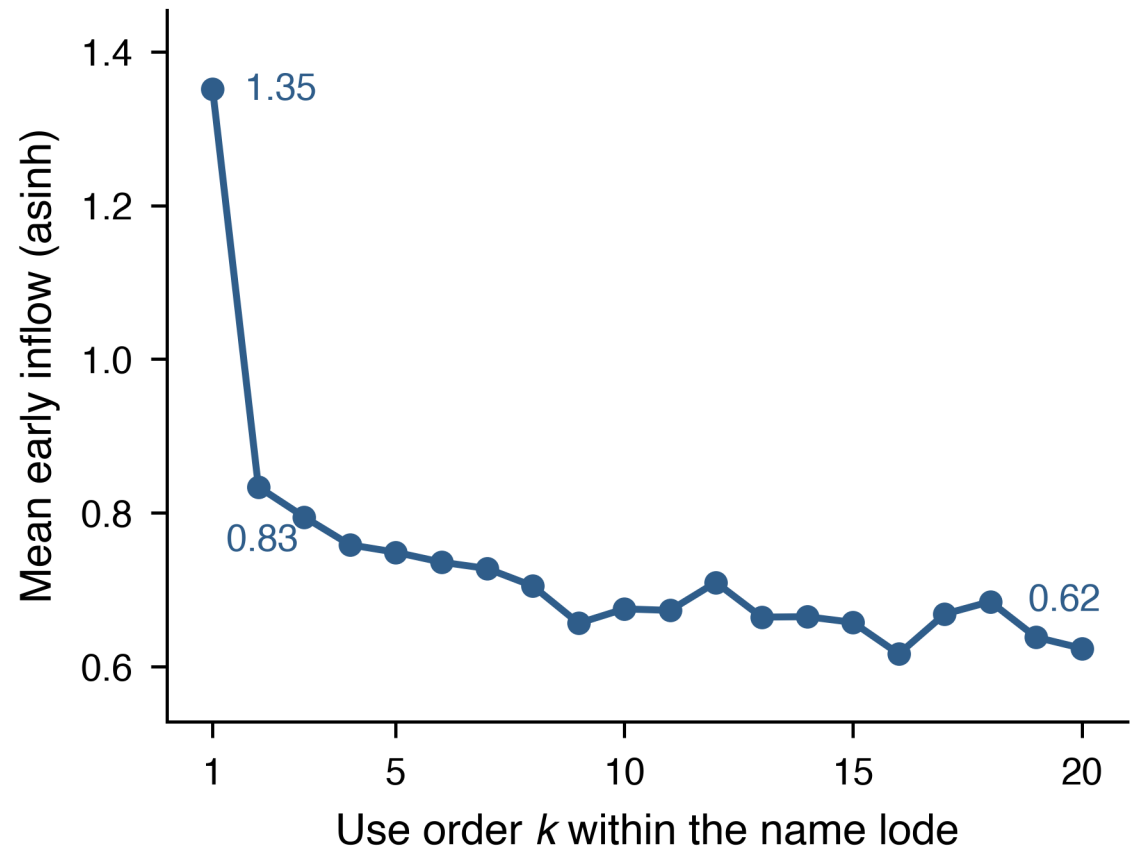

**Figure 2.** Early inflow under repeated use of the same name. The vertical axis shows the mean inverse hyperbolic sine (asinh) of the amount bought on the bonding curve (in BNB) within 60 minutes of creation, and the horizontal axis shows the token's use order k within its name lode, where $k = 1$ is the first use of the name and lodes are defined over all launches on the platform. The sample consists of the 140,686 default-template tokens with $k = 1$ to 20 in clean repeated lodes of the exploration sample, of which 32,756 have $k = 1$. Clean repeated lodes are lodes used at least twice whose normalized name is shorter than 20 characters and is not purely numeric or symbolic.

The coordination explanation further implies that names that draw capital are copied quickly. Panel B of Table 5 applies a discrete-time hazard model to the first tokens of 103,235 exploration names to estimate the risk that a name is first used by another creator address within 30 days. A one-unit increase in the first token's early inflow raises this hazard by 92.4%, and by 97.7% in the validation sample, whereas migration of the first token does not raise it ($p = 0.559$) and lowers it in the validation sample ($p = 0.025$). Among the 26,502 names copied within 30 days, the median copy arrives 1.1 minutes after the first token, and 67.5% arrive within 60 minutes. Defining copying by creator entity rather than by address also gives a median of 1.1 minutes, so rapid copying does not mainly reflect multi-address issuance by one identifiable entity.

**Table 5 Reuse and copying of names**

| | Exploration | Validation |
|---|---|---|
| **Panel A. Decay with use order within lodes (dependent variable is early inflow)** | | |
| Use order | −0.456*** | −0.449*** |
| | (0.030) | (0.052) |
| Use order × imitation weight | 0.009 | 0.018 |
| | (0.014) | (0.012) |
| Lode and creation-minute fixed effects | Yes | Yes |
| Observations | 69,796 | 66,907 |
| **Panel B. Hazard of first copying by another creator address** | | |
| Early inflow of first token | 0.654*** | 0.682*** |
| | (0.013) | (0.020) |
| First token migrated | −0.049 | −0.230** |
| | (0.084) | (0.102) |
| First tokens | 103,235 | 92,641 |
| Names copied within 30 days | 26,502 | 23,269 |
| Median copy arrival time (minutes) | 1.1 | 0.8 |
| Share of copies arriving within 60 minutes | 67.5% | 70.5% |
| Median arrival time, entity-based definition (minutes) | 1.1 | 0.8 |

Notes: Lodes and use order are defined over all 513,647 launches on the platform. Panel A uses the second and later uses of names in clean repeated lodes. Use order is the inverse hyperbolic sine of the token's order k within its lode. The imitation weight is the posterior probability, from a two-component Gaussian mixture model, that a repetition is an imitation. The model uses the time between consecutive uses and the normalized edit distance between the names as features, and its parameters are estimated on the exploration sample and applied to the validation sample. The regressions also control for the main effect of the imitation weight. In Panel B, the first token of a name is the earliest launch of its lode and uses the default template, and copies are counted among default-template launches. Panel B is estimated with a discrete-time hazard model with a complementary log-log link over a 30-day risk period, controlling for survival-day dummies (pooled from day 15 onward), and the exponential of a coefficient is a hazard ratio. Copy arrival time is the time between the creation of a name's first token and the first use of the name by another creator address. Standard errors clustered by creator entity are in parentheses. *, ** and *** denote significance at the 10%, 5% and 1% levels, respectively.

Rent-seeking theory distinguishes transfers of wealth from the real resources spent competing for them [21], and under particular contest designs and information conditions contestable rents can be fully dissipated [22]. Table 6 summarizes 30-day profits and losses for the 210,661 exploration tokens with at least one purchase, valuing positions still held at the end-of-window price. Losing outside buyers lose $143.86 million in total, winning outside buyers gain $136.20 million, and creators' own positions gain $7.02 million net. Because returns fall significantly with the order of entry (Table 4), these transfers favor earlier entrants. Creating the tokens cost $0.18 million on chain in gas and launch fees, a median of $0.0996 per token and 0.13% of the losses of outside buyers. Off-chain costs such as promotion are

not observed, so the ledger measures the size of the transfers but cannot identify the total real resource dissipation.

**Table 6 Profit-and-loss ledger of competition over names**

| Item | Amount (US$ million) | Relative to losses of outside buyers |
|---|---|---|
| Losses of outside buyers with a loss | 143.86 | 100% |
| Gains of outside buyers with a gain | 136.20 | 94.67% |
| Net gain on creators' own positions | 7.02 | 4.88% |
| On-chain cost of creating the tokens (gas and launch fees) | 0.18 | 0.13% |

Notes: The sample consists of the 210,661 exploration tokens with at least one purchase, and the window is the 30 days (720 hours) after creation. Profit or loss equals sale proceeds plus the value of the remaining position at the end-of-window price, minus the purchase cost, so it includes unrealized profit or loss on positions held at the end of the window. The net loss of outside buyers as a whole is $7.66 million. Creators' own positions gain $7.64 million and lose $0.62 million, and these amounts are transfers among buyers rather than resource costs. The on-chain cost is the gas paid for each token's creation transaction, read from its receipt, plus the launch fee recorded on chain at creation, and 999 of the tokens paid a launch fee. Off-chain costs such as promotion and labor are not observed and are not included.

## 4. Conclusion

On a launchpad where tokens carry no cash flows, names have a small within-creator effect on attracting outside buyers that does not replicate out of sample. Conditional on early inflow and within creator, their residual effects on migration, peak market capitalization and drawdown are statistically equivalent to zero, and buyers who follow appealing names earn no economically meaningful premium. A name's appeal declines with repeated use, names whose first token draws capital are copied within minutes, and the resulting rents become mainly transfers among buyers. Under the maintained assumptions, this evidence favors the view that names act as focal points whose value depends on the context of their use over the view that names convey information.

If names coordinate attention, name effects should be stronger where other information is scarcer. Where linguistic salience can be contested cheaply, as with app-store names and short-video hashtags, competition may likewise mainly redistribute wealth toward early movers. The evidence comes from five months of data on one platform. The appeal index is fitted to early inflow, the validation tests were run a second time after a data error was corrected, and the survival and migration results depend on the specification. These conclusions await tests on other platforms and over longer periods.

### CRediT authorship contribution statement

Dingding Cao: Conceptualization, Methodology, Software, Data curation, Formal analysis, Writing – original draft. Yujing Zhong: Investigation, Validation, Data curation. Han Wang: Investigation, Validation. Xian Pan: Resources, Visualization. Wei Yang and Rizwan Akhtar: Supervision, Project administration, Writing – review and editing.

### Declaration of competing interest

The authors declare no known competing financial interests or personal relationships that could have appeared to influence the work reported in this paper.

### Funding

This work was supported by the Baoshan University Doctoral Research Startup Fund Program 2025 (Grant No. BSKY2542) and the Joint Special Fund for Basic Research of Local Undergraduate Institutions in Yunnan Province (Grant No. 202101BA07001-200).

### Data and code availability

On-chain data are drawn from public blockchains and public data services. The appeal-index weights, the launch-template and creation-cost tables read from the chain, the estimation code, all test outputs and the records of both validation runs are available in the replication package at: https://doi.org/10.5281/zenodo.23063777. The census master table, the three-layer name features and the codebook are released with the companion study [10].

### Declaration on the use of large language models

During the preparation of this manuscript, large language models were used for language polishing, code translation and refinement, and assisted coding of the reference-layer features. All manuscript content was written by the authors, and full academic responsibility is assumed by the authors.

### Ethics statement

This research uses only public on-chain data and public APIs. On-chain data require no anonymization.